\documentclass[aip,jcp,reprint,citeautoscript,floatfix]{revtex4-1}
\usepackage{natbib}
\usepackage{graphicx}
\usepackage[caption=false,lofdepth,lotdepth]{subfig}
\usepackage{import}
\usepackage{lmodern}
\usepackage[colorinlistoftodos,color=black!30]{todonotes}
\usepackage{amsmath, amsthm, amssymb, mathtools}
\usepackage[T1]{fontenc}
\usepackage{microtype}
\usepackage[utf8]{inputenc}
\usepackage{url}
\usepackage{enumerate}
\usepackage{physics} 
\usepackage{bm} 
\usepackage{booktabs,tabularx}
\usepackage{cleveref} 
\usepackage{accents}
\usepackage{layouts}
\usepackage{booktabs}
\usepackage{float}
\crefname{equation}{Eq.}{Eqs.}
\crefname{section}{Sec.}{Secs.}
\crefname{table}{Tab.}{Tabs.}
\crefname{figure}{Fig.}{Figs.}
\crefname{subfigure}{Fig.}{Figs.}

\usepackage[nolist,nohyperlinks]{acronym}
\usepackage{pgfplots}
\usetikzlibrary{pgfplots.groupplots}
\pgfplotsset{compat=1.5}
\usepackage{siunitx} 
\usepackage{multirow} 
\usepackage{bbm}
\graphicspath{{figures/}}

\begin{document}
\definecolor{plt_blue}{RGB}{31,119,180}
\definecolor{plt_orange}{RGB}{255,127,14}
\definecolor{plt_green}{RGB}{44,160,44}
\definecolor{plt_red}{RGB}{214,39,40}

\title{
Generalized Path Reweighting 
 and
 History-Dependent
 Free Energies
}
\author{Titus S. van Erp}
\email{titus.van.erp@ntnu.no}
\affiliation{Norwegian University of Science and Technology, Department of Chemistry and Biomedical Science,  NO-7491 Trondheim, Norway}
\author{Daniel T. Zhang}
\affiliation{Norwegian University of Science and Technology, Department of Chemistry and Biomedical Science,  NO-7491 Trondheim, Norway}
\affiliation{Research Institute for Interdisciplinary Science, Okayama University, Okayama 700-8530, Japan}
\author{Elias Wils}
\affiliation{IBiTech - BioMMedA group, Ghent University, Corneel Heymanslaan 10, entrance 97, 
9000 Gent, Belgium}
\author{Sina Safaei}
\affiliation{IBiTech - BioMMedA group, Ghent University, Corneel Heymanslaan 10, entrance 97, 
9000 Gent, Belgium}
\author{An Ghysels}
\affiliation{IBiTech - BioMMedA group, Ghent University, Corneel Heymanslaan 10, entrance 97, 
9000 Gent, Belgium}

\begin{abstract}
Transition interface sampling (TIS) and replica exchange TIS (RETIS) are powerful methods for computing rates of rare events inaccessible to straightforward molecular dynamics (MD) simulations.
Path reweighting extends their output, enabling the evaluation of diverse thermodynamic and kinetic quantities, including reaction prediction metrics, activation barriers, committor functions, and free energies. The recently developed 
$\infty$RETIS
algorithm boosts parallel efficiency through asynchronous replica exchanges in the infinite-swap limit, eliminating the wall-time bottlenecks of conventional RETIS. 
This approach introduces fractional samples and biased sampling distributions, requiring a generalized path reweighting framework, for which we derive expressions demonstrating how exact dynamic and thermodynamic variables can be computed.
We then focus on a special class of free energy surfaces defined by history-dependent conditions, 
whose values are influenced by
kinetic factors such as particle mass and friction, unlike standard unconditional free energy surfaces. Even with suboptimal reaction coordinates, these conditional free energies can reveal kinetically relevant barriers that may be misrepresented by standard unconditional free energies, thereby providing a rigorous and versatile tool for characterizing complex molecular transitions.
\end{abstract}

\maketitle
\section{Introduction}

\label{sec:introduction}
Transition path sampling (TPS), introduced in a seminal paper by Dellago and co-workers,
\cite{TPS,TPSReview1, TPSReview2} is a Monte Carlo (MC) method for efficiently sampling rare reactive trajectories that connect stable states, without altering the system’s dynamics or potential energy surface. In TPS, new trajectories are generated using path-generating MC moves such as \emph{shooting},\cite{shoot} in which a phase point is selected from an existing path and slightly perturbed, followed by forward and backward integration of the equations of motion 
via two short molecular dynamics (MD) runs 
to construct a new trial path. TPS enables analysis of the reaction mechanism, identification of potentially parallel reaction channels, and quantification of their relative probabilities. To compute absolute probabilities and reaction rates, a series of path ensembles must be sampled, as done in transition interface sampling (TIS)\cite{TIS}, and more efficiently in replica exchange TIS (RETIS)\cite{RETIS}.
 
Rogal et al.\cite{Rogal2010} demonstrated that the simulation output of TIS, RETIS, and related methods such as forward flux sampling (FFS)\cite{Allen2005} can be used to compute not only reaction rates but also a broad range of other thermodynamic and dynamical observables via path reweighting. This framework was later generalized in Ref.~\citenum{vanErp2016} to accommodate cases where the number of sampled trajectories differs between path ensembles, which was an important consideration for TIS and for the latest algorithmic improvement of RETIS, known as $\infty$RETIS.\cite{InfRET1, PNAS2024} In conventional RETIS, where ensembles are updated sequentially, the number of sampled paths per ensemble is typically equal, making the generalization less critical. However, when ensembles are run in parallel,
as is possible in TIS and is standard in $\infty$RETIS, differences in average path length naturally lead to imbalances, with shorter-path ensembles accumulating more samples than those with longer paths.

The same path-length imbalance explains why RETIS ensembles are typically not run in parallel. Frequent replica exchange moves between ensembles require synchronization, causing hardware groups (called \emph{workers}) handling short-path ensembles to wait for those working on longer paths before performing the next replica exchange move. In contrast, TIS can be parallelized efficiently, as each ensemble is sampled independently without such coordination. Thus, although RETIS is significantly more efficient in terms of CPU usage, TIS can still be competitive in wall-clock time when run on highly parallel computing architectures.
The $\infty$RETIS algorithm resolves the limitations of conventional RETIS by simultaneously optimizing both CPU efficiency and wall-time efficiency, thereby outperforming both RETIS and TIS on these metrics. We refer the reader to Refs.~\citenum{InfRET1,PNAS2024, WF} for a detailed description of the algorithm, as the present work focuses on the analysis of path data rather than on the specifics of how such data are generated.

To establish its excellent parallelizability,
 $\infty$RETIS uses multiple workers that perform shooting-type moves in parallel across different ensembles. Crucially, the number of workers is smaller than the number of ensembles, meaning some ensembles remain temporarily unoccupied. After each shooting move, the algorithm mimics the effect of an \emph{infinite number} of replica exchanges among all free ensembles via fractional samples that reflect the relative frequency with which a given path would reside in each ensemble during such an infinite-swapping round.
In addition, $\infty$RETIS is often combined with advanced shooting strategies, such as the \emph{wire-fencing} move with \emph{high-acceptance},\cite{WF} which enhances sampling efficiency and improves trajectory decorrelation. The high-acceptance protocol alters the sampling distribution, which must be corrected during analysis by weighting each sample inversely proportional to the corresponding high-acceptance weight. The fractional samples and the biased path distribution require a further generalization of the path reweighting scheme, and a detailed mathematical derivation of this generalization constitutes the first objective of our paper.

Path reweighting opens up a range of interesting possibilities, such as Arrhenius-based activation energies, reaction prediction metrics, and committor-related quantities, which are not discussed in this paper and will be reported in future work. In this paper, we  focus, as a second objective, on the analysis of a specific path observable, namely the history-dependent conditional free energy surface or profile, which can be computed from $\infty$RETIS output using path reweighting. Although conditional free energies have been reported previously (e.g.\ in Refs.~\citenum{PNAS2024,MinhCrystalNucleation2025,ZhangRETIS2025}), they have not yet received the broader attention they merit, despite their intriguing and distinctive characteristics.
A conditional free energy profile is obtained not by simply histogramming configuration points from a simulation sampling the equilibrium distribution, but by histogramming phase points along paths belonging to a specific path ensemble, for example, paths starting from $A$ or paths starting from 
$B$. Such profiles possess features absent in standard free energy profiles. Unlike the latter, they are not purely thermodynamic; their shape depends on the underlying kinetics. Consequently, the profile can change when particle masses are altered or when the friction coefficient in a Langevin dynamics simulation is varied, effects that leave plain free energies unchanged. Furthermore, conditional free energies can reveal kinetically relevant free energy barriers even when using a suboptimal reaction coordinate. In such cases, standard free energies typically underestimate the true barriers that the system must overcome, whereas the conditional profiles retain meaningful barrier information. Finally, conditional free energies obtained from different path ensembles or conditions can be combined to reconstruct the standard unconditional free energy profile, providing a comprehensive thermodynamic picture.

\section{path ensembles}
\label{Sec:pathens}

In path sampling, a path $X$ is depicted as a discrete sequence of phase points, called time slices $X = [x_0, x_1, \ldots x_L]$. 
Each time slice $x_l$ 
is a phase point
representing the coordinates and velocities of atoms at a specific time  
$t =l \Delta t$, where $L$ denotes the trajectory's length, and $\Delta t$ represents a small time step.  
The path probability
distribution equals
\begin{eqnarray}
\rho(X) = 
\rho(x_0) p(x_0 \rightarrow x_1) p(x_1 \rightarrow x_2) \ldots p(x_{L-1} \rightarrow x_L),
\label{eq:rhoX}
\end{eqnarray}
where $\rho(\cdot)$ is the equilibrium distribution of a phase point, which is conserved by the dynamics (e.g., the Boltzmann distribution in the case of canonical NVT dynamics), and $p(x \rightarrow y)$ is the probability that the system's dynamics produces $y$ after a $\Delta t$ time step from $x$. The index “0” in $x_0$ does not refer to a specially prepared initial configuration. 
Rather, it serves as a generic label for the first phase-point of the considered path $X$.

In the original TPS method, the path length $L$ was a user-defined fixed integer that was the same for all sampled paths. In contrast, in TIS the path length becomes a flexible quantity, as the start and end points are determined by the conditions of the respective path ensembles discussed below.
 
The goal of path sampling is to generate short MD trajectories according to the correct distribution, Eq.~\ref{eq:rhoX}, under specific constraints. While this could in principle be done by analyzing a long MD run, the MC approach efficiently focuses on rare, reactive trajectories near the barrier, avoiding time spent in stable states.
TPS,\cite{shoot} for example, samples such transitions by requiring all paths to start in $A$ and end in $B$, typically with fixed length $L$. However, to obtain quantitative dynamical information, a single transition path ensemble is not enough; instead, a series of ensembles is needed, as in TIS\cite{TIS} to cover the non-reactive paths as well.

The TIS path ensembles are defined using interfaces $\lambda_0, \lambda_1, \ldots, \lambda_n$, based on an order parameter $\lambda(x)$ that measures reaction progress.
 Each interface $\lambda_i$ corresponds to the hypersurface $\{x \mid \lambda(x) = \lambda_i\}$. The reactant state $A$ is defined by $\{x |\lambda(x) < \lambda_0 = \lambda_A\}$, and the product state $B$ by $\{x |\lambda(x) > \lambda_n = \lambda_B\}$.

TIS ensembles consist of paths with variable lengths $L$ that satisfy specific initial/final conditions and a minimal progress condition. For example, the $[i^+]$ ensemble contains all paths that start by crossing $\lambda_A$ toward the barrier region and end by either returning to $A$ or reaching $B$.
Additionally, each path must cross $\lambda_i$ at least once to belong to this ensemble.
In RETIS,\cite{RETIS} an additional $[0^-]$ ensemble is included, which explores dynamics confined to state $A$. 

In summary, a trajectory $X$ belongs to $[0^-]$ or $[i^+]$ if it satisfies the following conditions:
\begin{align}
\label{Eq:cond}
&X \in [0^-] \textrm{ if } 
\begin{cases}
\lambda(x_0) > \lambda_A, \\
\lambda(x_L) > \lambda_A, \textrm{ and} \\
\lambda(x_l) < \lambda_A \textrm{ for } 0 < l < L 
\end{cases}   \\
&X \in [i^+] \textrm{ if } 
\begin{cases}
\lambda(x_0) < \lambda_A, \\
\lambda(x_L) < \lambda_A \textrm{ or } \lambda(x_L) > \lambda_B,  \\
\lambda_A < \lambda(x_l) < \lambda_B \textrm{ for } 0 < l < L, \textrm{ and}\\
\lambda_{\rm max}(X)>\lambda_i
\end{cases}\nonumber
\end{align}
with
\begin{align}
\lambda_{\rm max}(X)={\rm max}(\lambda(x_0),
\lambda(x_1), \ldots, \lambda(x_L)
)
\end{align}
Based on these definitions, we can define the 
(unnormalized) path distributions for each ensemble as
\begin{align}
\rho_E(X) = \rho(X) \cdot {\mathbbm 1}_E(X),\label{eq:1E}
\end{align}
where ${\mathbbm 1}_E(X)$ equals 1 if the path adheres to the ensemble $E$ conditions (Eq.~\ref{Eq:cond}), and 0 otherwise.
To shorten the notation we will write 
$\rho_i$ for $\rho_{[i^+]}$ and 
$\rho_{\bar{0}}$ for $\rho_{[0^-]}$.

Based on this, the ensemble average of a path observable $O(X)$ in   
$[0^-]$ and $[i^+]$
can be formally defined as:
 \begin{align}
 \left \langle O \right \rangle_{\overline{0}} =\frac{\int  O(X) \rho_{\overline{0}}(X) {\mathcal D} X}{
\int  \rho_{\overline{0}}(X) {\mathcal D} X
}, \quad
\left \langle O \right \rangle_i =\frac{\int  O(X) \rho_i(X) {\mathcal D} X}{
\int  \rho_i(X) {\mathcal D} X
}
\quad
\label{eq:pathaverage}
\end{align}
The integration over path space implies a summation over all possible path lengths, followed by integration over the phase points: 
$\int \ldots {\mathcal D} X=
\sum_{L=3}^\infty \int \ldots  {\mathrm d}x_0 {\mathrm d}x_1 \ldots  {\mathrm d}x_L$. 

As integration over the entire path space is practically infeasible, path sampling utilizes a MC approach, where paths are sampled according to the correct distribution $\rho_i(X)$,
using moves such as the shooting move\cite{shoot} that satisfy detailed balance.
Hence, after 
$N_i$ MC moves in ensemble $[i^+]$, the ensemble average 
is estimated as:
\begin{align}
\left \langle O \right \rangle_i 
 = \frac{1}{N_i} \sum_{ j=1}^{N_i} O(X_{{\rm sam }j}^{(i)}) \label{Eq:av0}
\end{align}
where $X_{{\rm sam} j}^{(i)}$ is the $j$th sample in the MC Markov chain that resulted form the simulation 
of path ensemble $[i^+]$.

In the MC Markov chain, the same path can be sampled multiple times since MC requires that the old state is resampled again after a rejected  move. In addition, since RETIS applies replica exchange moves between ensembles, the same path can also reappear in the same ensemble after two or more replica exchange moves. Let $m_{ji}$ 
be the multiplicity, i.e.\ how often the $j$th sample is resampled in the $[i^+]$ 
path ensemble simulation
before it is pushed out of the MC Markov chain, e.g., due to a successful shooting move applied on this path that got accepted and replaced it.
Eq.~\ref{Eq:av0} is then equivalent to
\begin{align}
\left \langle O \right \rangle_i 
 = \frac{\sum_{ j}^{N_i^{u}} m_{ji} O(X_{j}^{(i)})}{
\sum_{ j}^{N_i^u} m_{ji}
 } \label{Eq:av1}
\end{align}
where $N_i^u \leq N_i$ are the number of unique samples $X_{j}^{(i)}$ in ensemble $[i^+]$, with $\sum_j^{N_i^u} m_{ji} = N_i$.
The same applies to the $[0^-]$ ensemble.

In the $\infty$RETIS method,\cite{PNAS2024, InfRET1} the use of fractional samples and weight factors results in a mathematical expression that is somewhat more complex than Eq.~\ref{Eq:av1}.
After each shooting move, the effect of an \emph{infinite number} of replica exchanges between all free ensembles is mimicked by calculating the fraction of times a given path would be sampled in a particular ensemble if an infinite number of swaps were performed.
The cumulative sum of these fractions over multiple rounds of infinite swapping, in which each unique sample participates while alive in the Markov chain, is denoted by \(\mu_{ji}\). This quantity represents the non-integer multiplicity 
with which the $j$th
$\infty$RETIS path $X_j$ is sampled in ensemble $[i^+]$,
analogous to the integer counts \(m_{ji}\) in Eq.~\ref{Eq:av1}.
Note that in $\infty$RETIS a single universal index $j$ is used to label paths, and the ensemble superscript $(i)$ is omitted (i.e., we write $X_j$ instead of $X_j^{(i)}$, which would denote the $j$th path of ensemble $i$). The rationale for this notation is explained at the end of this section.
The sum of the fractional multiplicities gives the total fractional number of trajectories $\eta_i = \sum_j \mu_{ji}$ in ensemble $[i^+]$, similarly to $N_i$ in Eq.~\ref{Eq:av1}.

Additionally, $\infty$RETIS 
standardly
employs advanced shooting moves such as wire-fencing\cite{WF}, which utilize a biased sampling distribution in which $\rho_i(X)$ is replaced by $\rho_i(X) w_i(X)$, with $w_i(X)$ representing the high-acceptance weight of path $X$ in ensemble $[i^+]$ (or $w_{\bar{0}}$ for ensemble $[0^-]$).
This adjustment in the sampling distribution optimizes the acceptance rate of the MC moves. However, it necessitates weighting each sampled path by the inverse of $w_i$ to ensure unbiased ensemble averages. Consequently, Eq.~\ref{Eq:av1} can be reformulated as:
\begin{align}
\left \langle O \right \rangle_i 
 = \frac{\sum_{ j}  \mu_{ji} O(X_{j})/w_{i}(X_j)
 }{
\sum_{ j}  \mu_{ji}/w_{i}(X_j)
 } \label{Eq:av2}
\end{align}
where $j$ now runs over all sampled trajectories of the full
$\infty$RETIS
simulation, without being restricted to a specific path ensemble.
This indicates a subtle shift in perspective on how path ensemble averages are viewed: the focus moves from ensembles containing paths to paths themselves, accompanied by a corresponding frequency list indicating their visitation across multiple ensembles.
This alternate viewpoint better captures the unique nature of $\infty$RETIS, which is most naturally understood by focusing on individual paths rather than ensembles, since each path is sampled across many ensembles. Consequently, the main output of the $\infty$RETIS method is a record for each path, containing a list of path observables along with the $\mu_{ji}$ and $w_i(X_j)$ values.

\section{Path and phase space ensemble averages}
\label{sec:PathPhaseAv}

All kinetic and thermodynamic observables of interest derived from $\infty$RETIS output data can be expressed either as $[0^+]$ path-ensemble averages or as combinations of the $[0^+]$ and $[0^-]$ ensemble averages.
The ensembles $[i^+]$ with $i>0$ are therefore primarily used to improve the statistical accuracy of $\langle \ldots \rangle_0$ path averages, as $\langle \ldots \rangle_i$ does not, by itself, correspond to a commonly studied physical observable. This is further discussed in Sec.~\ref{Sec:PathRew}.

For example, an important quantity is the crossing probability
\(P_A(\lambda|\lambda_A)\), i.e., the probability that an exit from state \(A\) by crossing \(\lambda_A\) reaches at least \(\lambda\) before returning to \(A\), which can be written as
\begin{align}
P_A(\lambda \mid \lambda_A) &=
\left\langle
\theta\!\left(\lambda_{\mathrm{max}}(X) - \lambda\right)
\right\rangle_0 
\label{eq:Plam}
\end{align}
where $\theta$ refers to the Heaviside function.
The rate then follows as \(k_{AB} = f_A P_A(\lambda_B \mid \lambda_A)\), where the flux \(f_A\) can be obtained\cite{RETIS} from the average path lengths \(\langle L(X) \rangle_0\) and \(\langle L(X) \rangle_{\bar 0}\). 

Beyond path averages, where each trajectory contributes a single value, $\infty$RETIS also enables the evaluation of phase-space or configuration-space averages 
by accumulating contributions from all time slices along the sampled paths.
In this procedure, the initial and final points, $x_0$ and $x_L$, are excluded so that all contributing time slices belong to the same order parameter region: $\lambda(x) < \lambda_A$ for $[0^-]$ and $\lambda_A < \lambda(x) < \lambda_B$ for $[i^+]$.
The resulting set of phase points therefore consists of configurations that  
reside in the stable state $A$ or lie outside $A$ but belong to trajectories originating from it.
Consequently, the phase- or configuration-space ensemble averages accessible from an $\infty$RETIS simulation are restricted to the \emph{overall state} $\mathcal{A}$: phase points inside $A$ or having visited $A$ more recently than $B$.\cite{TIS}

Let us first consider a phase-space average of an observable $o(x)$ restricted to state $A$, which can conceptually be obtained from a very long MD simulation by sampling all phase points $x$ for which $\lambda(x) < \lambda_A$, while suspending the collection whenever $\lambda(x) > \lambda_A$ and resuming it only after the system has returned below $\lambda_A$. This procedure is equivalent to extracting trajectory segments from the full MD trajectory such that each segment satisfies the definition of the path ensemble $[0^-]$, Eq.~\ref{Eq:cond}, and collecting all contributions of the phase points along these segments. $\infty$RETIS generates trajectories distributed according to the same path ensemble, i.e., statistically identical to the ensemble obtained by cutting such segments from an infinitely long MD trajectory. Therefore, the phase-space ensemble average can be rewritten as a path-ensemble average by introducing the following path functional,
\begin{align}
O(X;o) = \sum_{l=1}^{L(X)-1} o(x_l \mid X) .
\end{align}
As a result, the phase-space average of $o(x)$ restricted to state $A$ can be formulated as a path-space average of
$O(X;o)$ within the $[0^-]$ ensemble:
\begin{align}
    \left\langle o(x) \right\rangle_{A}
    &=
    \frac{
        \left\langle O(X;o) \right\rangle_{\bar{0}}
    }{
        \left\langle 
        \hat{L}(X) \right\rangle_{\bar{0}}
    } ,
    \label{eq:phase_avg_A}
\end{align}
where $\hat{L}(X)=L(X)-1$ is the number of time slices of the trimmed path
$\hat{X}=(x_1, x_2, \ldots, x_{L-1})$, obtained by removing the two outer points of $X$. The division by $\left\langle \hat{L}(X) \right\rangle_{\bar{0}}$ ensures proper normalization per phase point rather than per path.

By including the $[0^+]$ ensemble, the phase-space average can be extended from
$A$ to the overall state $\mathcal A$ as
\begin{align}
    \left\langle o(x) \right\rangle_{\mathcal A}
    &=
    \frac{
        \left\langle O(X; o) \right\rangle_{\bar{0}}
        +
        \left\langle O(X; o) \right\rangle_{0}
    }{
        \left\langle \hat{L}(X) \right\rangle_{\bar{0}}
        +
        \left\langle \hat{L}(X) \right\rangle_{0}
    } ,
    \label{eq:phase_av_overall}
\end{align}
Here, we have used the fact that an infinitely long MD trajectory generates an equal number of path segments in the $[0^-]$ and $[0^+]$ ensembles. For unbounded systems, such as membrane systems, an additional $\lambda_{-1}$ interface may be required, which invalidates this assumption. In such cases, a correction to Eq.~\ref{eq:phase_av_overall} is needed, as discussed in Ref.~\citenum{ghysels2021exact} and detailed in Appendix~\ref{sec:cor_endR}. 

In the same work~\cite{ghysels2021exact}, it was shown that Eq.~\ref{eq:phase_av_overall}, 
when expressed in terms of phase-space partition integrals,
\begin{align}
\left\langle o(x) \right\rangle_{\mathcal A} 
= \frac{\int \rho(x)\, o(x)\, \mathbbm{1}_{\mathcal A}(x)\, \mathrm{d}x}{\int \rho(x)\, \mathbbm{1}_{\mathcal A}(x)\, \mathrm{d}x},
\label{eq:integ_expression}
\end{align}
requires $x$ to be interpreted as an \emph{extended} or \emph{trajectory} phase point that includes not only positions and momenta but also the relevant stochastic noise variables. This is necessary because $\mathbbm{1}_{\mathcal A}(\cdot)$ is a history-dependent function and can therefore depend on these noise variables when the dynamics are non-deterministic.

This extended phase-space formalism
ensures that the integral,
Eq.~\ref{eq:integ_expression},  is well defined and provides a basis for expressing unconditional ensemble averages in terms of conditional ones:
\begin{align}
\left\langle o(x) \right\rangle
&= \frac{\int \rho(x)\, o(x)\, \, \mathrm{d}x}{\int \rho(x)\,  \mathrm{d}x} \nonumber \\
&= \frac{\int \rho(x)\, o(x)\, [\mathbbm{1}_{\mathcal A}(x) 
+\mathbbm{1}_{\mathcal B}(x)
]\, \mathrm{d}x}{\int \rho(x)\,  \mathrm{d}x}
\nonumber \\
&= \frac{\int \rho(x)\, o(x)\, \mathbbm{1}_{\mathcal A}(x) 
\, \mathrm{d}x}{\int \rho(x)\, \mathbbm{1}_{\mathcal A}(x) \,  \mathrm{d}x}
\frac{\int \rho(x)\,  \mathbbm{1}_{\mathcal A}(x) 
\, \mathrm{d}x}{\int \rho(x)\,   \mathrm{d}x}
\nonumber \\
&+\frac{\int \rho(x)\, o(x)\, \mathbbm{1}_{\mathcal B}(x) 
\, \mathrm{d}x}{\int \rho(x)\, \mathbbm{1}_{\mathcal B}(x) \,  \mathrm{d}x}
\frac{\int \rho(x)\,  \mathbbm{1}_{\mathcal B}(x) 
\, \mathrm{d}x}{\int \rho(x)\,   \mathrm{d}x}
\nonumber \\
&= P_{\mathcal A} \left\langle o(x) \right\rangle_{\mathcal A}+
P_{\mathcal B} \left\langle o(x) \right\rangle_{\mathcal B}
\end{align}
where $P_{\mathcal A}$ and $P_{\mathcal B}$ are the probabilities of states $\mathcal A$ and $\mathcal B$, satisfying $P_{\mathcal A} + P_{\mathcal B} = 1$, and related via the forward and backward rates as
\[
P_{\mathcal A} k_{AB} = P_{\mathcal B} k_{BA}.
\]
This relation follows from equilibrium, where the net flux between $\mathcal A$ and $\mathcal B$ vanishes, implying that the forward and backward reactive fluxes are equal.

Thus, availability of $\infty$RETIS results for both the forward and backward processes allows direct evaluation of unconditional ensemble averages via
\begin{align}
\left\langle o(x) \right\rangle
&= \frac{k_{BA}
\left\langle o(x) \right\rangle_{\mathcal A}+k_{AB} 
\left\langle o(x) \right\rangle_{\mathcal B}
}{k_{AB}+k_{BA}} 
\end{align}

\section{Path reweighting}\label{Sec:PathRew}
As discussed in Sec.~\ref{sec:PathPhaseAv}, the two key path-ensemble averages are $\langle \cdots \rangle_{\bar{0}}$ and $\langle \cdots \rangle_{0}$, which are naturally associated with the $[0^-]$ and $[0^+]$ ensembles, respectively. The latter, however, often requires the inclusion of path ensembles $[i^+]$ with $i>0$ in order to enhance sampling in the barrier region and to obtain reliable estimates of rare-event probabilities that are too small to be observed directly in the $[0^+]$ ensemble within feasible simulation times.
In contrast, sampling of $\langle \cdots \rangle_{\bar{0}}$ is comparatively straightforward, as it covers the high-probability region of the 
reactant state.
Path reweighting therefore provides a systematic means to improve the statistics of the $\langle \cdots \rangle_{0}$ ensemble averages by explicitly incorporating simulation data from the $[i^+]$ ensembles in addition to the $[0^+]$ ensemble itself.

Remarkably, despite the complexity arising from several factors (such as choosing optimal weights across ensembles to minimize error, avoiding overcounting in overlapping ensembles, and accounting for fractional sampling and high-acceptance weights), the final result reduces to assigning a single weight to each path.
Once these path weights are determined, a wide range of path-ensemble averages can be evaluated without further concern for ensemble distinctions or reweighting procedures. Specifically, the path-ensemble average of an observable $O$ can be written as
\begin{align}
\left\langle O \right\rangle_{\bar{0}} &= \sum_j \bar{\Lambda}_j O(X_j), \qquad
\left\langle O \right\rangle_{0} = \sum_j \Lambda_j O(X_j),
\label{eq:barLamLam}
\end{align}
where the sums run over all sampled paths, with weights defined such that $\bar{\Lambda}_j = 0$ for $X_j \notin [0^-]$ and $\Lambda_j = 0$ for $X_j \in [0^-]$, due to the fact that the $[0^-]$ ensemble does not overlap with any of the other ensembles.
Consequently, only one weight per path needs to be evaluated.
The weights are normalized, with $\sum_j \bar{\Lambda}_j=\sum_j \Lambda_j=1$.

The main output of $\infty$RETIS includes several key values for each sampled path $X_j$: the cumulative fractional samples $\mu_{jk}$ and the high-acceptance weights $w_k(X_j)$ for $k=\bar{0}, 0, 1, \ldots, n-1$, the path length $L(X_j)$, the maximum order parameter $\lambda_{\rm max}(X_j)$, and potentially a list of precomputed path observables. Additionally, each path is accompanied by a label indicating the file where the 
trajectory is saved, such that a variety of path observables not already provided can be computed afterwards.

To determine the effective weight factors $\bar{\Lambda}_j$ or $\Lambda_j$ for each sampled path $X_j$, the following procedure is used. First, the total (fractional) number of paths sampled in each ensemble $[k^+]$ is first computed as:
\begin{align}
    \eta_k = \sum_{j} \mu_{jk} \label{eq:etak}
\end{align}
for $k=\bar{0}, 0, 1, \ldots, n-1$.
Subsequently, the unbiased sampling weight $t_{jk}$ for each path $X_j$ and for each ensemble $k$ is calculated with appropriate scaling, ensuring that the sum of all unbiased weights in ensemble $k$ equals the total number of paths, $\eta_k$, as given by Eq.~\ref{eq:etak}:
\begin{align}
 &t_{jk} \propto \mu_{jk}/w_{k}(X_j) \nonumber \\
&\sum_{ j}  t_{jk} = {\eta}_k 
 \label{eq:tjk}
\end{align} 
This rescaling is permissible because path-ensemble averages (e.g., Eq.~\ref{Eq:av2}) are invariant under a global rescaling of the weights \(w_k\), such that \(t_{jk}\) may be rescaled arbitrarily. Enforcing \(\sum_j t_{jk}=\eta_k\) enables a direct comparison with the standard integer multiplicities \(m_{jk}\) of Eq.~\ref{Eq:av1}, which in Ref.~\citenum{vanErp2016} are proportional to each ensemble’s contribution to the total multi-ensemble path weight, thereby allowing the derivation of effective weights in \(\infty\)RETIS to proceed in a formally analogous manner upon replacing \(N_k\) and \(m_{jk}\) by \(\eta_k\) and \(t_{jk}\), respectively.

The weights $\bar{\Lambda}_j$ follow from the $k=\bar{0}$ values via
\begin{align}
\bar{\Lambda}_j = t_{j\bar{0}}/\eta_{\bar{0}} .
\label{eq:barLw}
\end{align}
Substitution of Eq.~\ref{eq:barLw} into Eq.~\ref{eq:barLamLam} is
equivalent to Eq.~\ref{Eq:av1} for $i=\bar{0}$, since it is restricted to a single
ensemble, $[0^-]$. As a consequence, there are no fractional samples arising from infinite exchange rounds with other ensembles (hence $\mu_{j\bar{0}} = m_{j\bar{0}}$). Moreover, no high-acceptance wire-fencing protocol is typically applied in either $[0^-]$ or $[0^+]$, such that $w_{\bar{0}}(X_j) = 1$ whenever $X_j \in [0^-]$. This results in $t_{j\bar{0}} = m_{j\bar{0}}$ and $\eta_{\bar{0}} = \sum_j m_{j\bar{0}}=N_{\bar{0}}$.

The $\Lambda_j$ weights should not solely rely on paths sampled in the $[0^+]$ ensemble, as this provides insufficient information for rare events. For instance, the overall crossing probability $P_A(\lambda_B \mid \lambda_A)$ could, in principle, be estimated from $[0^+]$ trajectories using Eq.~\ref{eq:Plam}, which reduces to the fraction of paths reaching state $B$. For rare events, however, this probability is typically so small that none of the sampled trajectories in $[0^+]$ reach $B$. The other $[i^+]$ ensembles are therefore essential, as they contain trajectories reaching at least $\lambda_i$, making further progression along the reaction coordinate more likely.

To combine information from the different ensembles for the $\Lambda_j$ computation, we derive expressions based on the Weighted Histogram Analysis Method (WHAM)\cite{WHAM1, WHAM2, WHAM3}, similar to Refs.~\citenum{Rogal2010} and \citenum{vanErp2016},
but tailored to $\infty$RETIS output. 
The remainder of this section presents the practical procedure for computing the $\Lambda_j$ weights, while theoretical justification and detailed derivations are given in Appendix~\ref{AppendixWHAM}.

A key step in this procedure is to assign, for each sampled path $X_j$, an integer index $K(X_j)$ that identifies the largest ensemble index for which $X_j \in [K^+]$:
\begin{align}
K(X)=\begin{cases}
k & \text{if } \lambda_{\rm max}(X) < \lambda_{n-1} \\
  & \quad \text{and } \lambda_k < \lambda_{\rm max}(X) \leq \lambda_{k+1}, \\
n-1 & \text{if } \lambda_{\rm max}(X) > \lambda_{n-1}.
\end{cases} \label{eq:K}
\end{align}
For convenience, we denote $K_j = K(X_j)$.
Following Ref.~\citenum{vanErp2016}, we then introduce
\begin{align}
Q_i=\left( 
\sum_{k=0}^i \eta_k [P_A(\lambda_k|\lambda_0)]^{-1}
\right)^{-1}
 \label{eq:Q}
\end{align} 
which leads to the expression for the path weights:
\begin{align}
\Lambda_{j} =
 Q_{K_j} \left( \sum_{k=0}^{n-1}
t_{jk} \right)
 \label{eq:Lamdef}
\end{align}

The overall weight factors $\Lambda_j$ in Eq.~\ref{eq:Lamdef} depend on the crossing probabilities $P_A(\lambda_i | \lambda_0)$ for $i = 1, 2, \ldots, n-1$ through $Q_K$ in Eq.~\ref{eq:Q}. Consequently, these probabilities must be determined first. At first glance, Eq.~\ref{eq:Plam} suggests that one could simply use Eq.~\ref{eq:barLamLam} with $O(X) = \theta(\lambda_{\rm max}(X) - \lambda_i)$. 
However, this would be circular, since it requires prior knowledge of $\Lambda_j$.

To resolve this circularity, we note that ensembles $[k^+]$ with $k \geq i$ contain only trajectories with $\lambda_{\rm max} > \lambda_i$, and thus provide no additional information for improving the statistical accuracy of the probability to reach $\lambda_i$. Their simulation results can therefore be omitted without loss of accuracy. By using only the results from ensembles with $k < i$, we obtain the following recursive relation for $P_A(\lambda_i \mid \lambda_0)$, as derived in Appendix~\ref{app:B} (for $i>0$):
\begin{align}
    P_A(\lambda_i | \lambda_0) &= Q_{i-1} \sum_{k=0}^{i-1} \eta_k \left \langle 
    \theta(\lambda_{\rm max}(X) - \lambda_i)
    \right \rangle_k \nonumber \\
    &= Q_{i-1} \sum_{k=0}^{i-1} \eta_k(i)
    \label{eq:pp}
\end{align}
where $\eta_k(i)$ denotes the sum of the unbiased sampling weights in $[k^+]$ that cross $\lambda_i$:
\begin{align}
    \eta_k(i)=
    \sum_j t_{jk} \, \theta\!\left(\lambda_{\rm max}(X_j) - \lambda_i\right)
    \label{eq:eta_brackets}
\end{align}

Since $P_A(\lambda_0 \mid \lambda_0) = 1$, it follows from Eq.~\ref{eq:Q} that $Q_0 = 1/\eta_0$.
We then evaluate $P_A(\lambda_i \mid \lambda_0)$ by forward iteration of Eqs.~\ref{eq:pp} and~\ref{eq:Q}, starting from $i=1$ and proceeding up to $i=n$.
The first few steps of this iteration can be written as:
\begin{align}
    P_A(\lambda_1 | \lambda_0) &= \frac{\eta_0(1)}{\eta_0},
\\ 
    P_A(\lambda_2 | \lambda_0)
    &=
    \frac{\eta_0(2)+\eta_1(2)}{\eta_0+\eta_1 [P_A(\lambda_1 | \lambda_0)]^{-1}}, 
    \nonumber \\
    P_A(\lambda_3 | \lambda_0)
    &=
    \frac{\eta_0(3)+\eta_1(3)+
    \eta_2(3)
    }{\eta_0+\eta_1 [P_A(\lambda_1 | \lambda_0)]^{-1}
    +\eta_2 [P_A(\lambda_2 | \lambda_0)]^{-1}
    }, \nonumber 
\end{align}

Hence, a single forward iteration over $i = 0, 1, 2, \dots$ yields the WHAM-weighted crossing probabilities $P_A(\lambda_i \mid \lambda_0)$ and the associated $Q_i$ factors of Eq.~\ref{eq:Q}, which define the path weights in Eq.~\ref{eq:Lamdef}. These weights allow observables in the $[0^+]$ ensemble to be computed with higher precision than from the raw $[0^+]$ data alone.

\section{Computing phase space/configuration space averages
and conditional free energies}
We previously showed that a phase-space ensemble average can be expressed in terms of path-ensemble averages via Eq.~\ref{eq:phase_av_overall}. This path average can then be evaluated using Eq.~\ref{eq:barLamLam}, with weights $\bar{\Lambda}_j$ and $\Lambda_j$ given by Eqs.~\ref{eq:barLw} and~\ref{eq:Lamdef}.
However, for the computation of phase-space or configuration-space averages, it is conceptually simpler, though mathematically equivalent, to reformulate the procedure as a weighted average over phase points rather than over paths. This exploits the fact that an infinitely long MD trajectory generates equal numbers of path segments in the $[0^+]$ and $[0^-]$ ensembles and that histogramming phase points assigns the same sample count to all points within a given path.

An effective weight can therefore be assigned to each phase point by collecting the time slices $x_1, x_2, \dots, x_{L-1}$ from each path and assigning each slice the weight of its parent path:
\begin{align}
\Lambda^*(x) =
\begin{cases}
\bar{\Lambda}_j, &
\text{if } x \in \hat{X}_j \textrm{ and }  X_j \in [0^-], \\ 
 \Lambda_j, & \text{if } x \in \hat{X}_j \textrm{ and } X_j \notin [0^-], 
\end{cases}
\end{align}
where $\hat{X}_j$ denotes the set of phase points along path $X_j$ excluding its end points $x_0$ and $x_L$. 
Once $\Lambda^*(x)$ is determined for each phase point, ensemble averages follow directly from a weighted sum over all phase points, independent of the path or ensemble from which they originate. However, whereas the sums of $\Lambda_j$ and $\bar{\Lambda}_j$ over all $j$ are unity, $\Lambda^*(x)$ is not normalized. Phase-space ensemble averages are therefore obtained as
\begin{align}
\left\langle o(x) \right\rangle_{\mathcal A} =
\frac{\sum_x \Lambda^*(x)\, o(x)}{
\sum_x \Lambda^*(x)
}. \label{eq:ostar}
\end{align}
where the sum runs over all phase points $x$ of all sampled  $\infty$RETIS trajectories $X_j$.

In the following, we focus on a particular phase-space ensemble average, namely the conditional free energy, which is defined as the logarithm of the probability distribution conditioned on either $\mathcal A$ or $\mathcal B$.

Let $\mathrm{\bf CV}$ denote a set of $m$ collective variables,
\[
\mathrm{\bf CV}(x) = \bigl(\mathrm{CV}_1(x), \mathrm{CV}_2(x), \ldots, \mathrm{CV}_m(x)\bigr),
\]
where each collective variable is a scalar function of the phase point $x$. In most practical applications, the collective variables depend only on the configuration and not on the momenta. Examples of CVs could be the angle of a permeating molecule, the end-to-end distance of a protein, or the velocity (does depend on momenta).

We define the multidimensional delta function
\[
\delta\!\left(\mathrm{\bf CV'} - \mathrm{\bf CV}(x)\right)
\equiv \prod_{i=1}^m \delta\!\left(\mathrm{CV}_i' - \mathrm{CV}_i(x)\right).
\]
Then, based on Eq.~\ref{eq:ostar}, the conditional (unnormalized) distribution in ensemble $\mathcal A$ is given by
\begin{align}
\varrho_{\mathcal A}(\mathrm{\bf CV'}) &=
\left\langle
\delta\!\left(\mathrm{\bf CV'} - \mathrm{\bf CV}(x)\right)
\right\rangle_{\mathcal A}
\nonumber \\
&\propto
\sum_x \Lambda^*(x)\,
\delta\!\left(\mathrm{\bf CV'} - \mathrm{\bf CV}(x)\right),
\label{eq:propA}
\end{align}
with the corresponding conditional free energy defined as
\begin{align}
F_{\mathcal A}(\mathrm{\bf CV'}) =
- k_B T \ln \left( \varrho_{\mathcal A}(\mathrm{\bf CV'})/\varrho_{\rm ref} \right).
\label{eq:FA}
\end{align}
where $\varrho_{\rm ref}$ is an arbitrary reference density that renders the argument of the logarithm dimensionless and sets the zero level of the free energy surface. Choosing $\varrho_{\rm ref} = \min \varrho_{\mathcal A}$ ensures that the minimum of the conditional free energy is zero, while free-energy differences along the CV remain independent of the particular choice of $\varrho_{\rm ref}$. Hence, the normalization by dividing by $\sum_x \Lambda^*(x)$, as in Eq.~\ref{eq:ostar}, is, in principle, unnecessary for Eq.~\ref{eq:propA}.

Moreover, if the $B \rightarrow A$ transition is also considered, the conditional free energy $F_{\mathcal B}(\mathrm{\bf CV})$ can be computed, followed by the determination of the corresponding unconditional free energy $F(\mathrm{\bf CV})$. 
In that case, one or both of the histograms, $\varrho_{\mathcal A}(\mathrm{\bf CV})$ and $\varrho_{\mathcal B}(\mathrm{\bf CV})$, must be rescaled to satisfy:
\begin{align}
\frac{
\int \varrho_{\mathcal B}({\rm \bf CV}) \, {\mathrm d}{\rm \bf CV}
}{
\int \varrho_{\mathcal A}({\rm \bf CV}) \, {\mathrm d}{\rm \bf CV}
} = \frac{k_{AB}}{k_{BA}}
\end{align}
After this rescaling, the total density becomes
\begin{align}
\label{eq:scaling}
\varrho({\rm \bf CV}) = \varrho_{\mathcal A}({\rm \bf CV}) + \varrho_{\mathcal B}({\rm \bf CV})
\end{align}
the corresponding unconditional free energy is
\begin{align}
\label{eq:free}
F({\rm \bf CV}) &= - k_B T \ln \varrho({\rm \bf CV}) \nonumber
\\
&=- k_B T \ln
\Big[ e^{-\beta F_{\mathcal A}({\rm \bf CV})} + e^{-\beta F_{\mathcal B}({\rm \bf CV})} \Big].
\end{align}

\section{Revealing Hidden Barriers via Conditional Free Energies}

\begin{figure}[ht!]
    \centering
\includegraphics[width=0.85\linewidth]{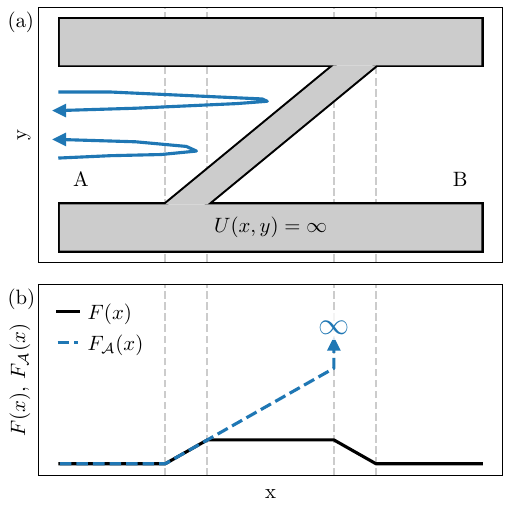}
    \caption{
(a) 2D representation of a potential energy surface showing a (blocked) “reaction tube,” where paths coming from A (depicted in blue) are unable to cross past the diagonal gray regions representing impenetrable walls to reach B. (b) The corresponding unconditional and conditional free energy profiles ($F(x)$, $F_\mathcal{A}(x)$) along $x$. Only $F_\mathcal{A}$ reveals the presence of an infinite barrier.
    }
    \label{fig:infwall}
\end{figure}

Free energy surfaces (FES) are very useful for obtaining qualitative insight into how a transition or reaction proceeds. However, any FES is necessarily a projection onto a reduced set of coordinates and can therefore be misleading. 
For this reason, we focus on conditional FES, which can, in certain cases, provide a more faithful visual representation of the actual transition process.
This is illustrated using the two-dimensional potential energy surface shown in Fig.~\ref{fig:infwall}, %
where the gray regions correspond to infinitely high, impenetrable walls.
The standard (unconditional) free energy profile along $x$ nevertheless displays only a modest barrier, since the probability density at a given value of $x$ is only weakly affected by the presence of a tilted wall in the orthogonal direction. In contrast, the conditional free energy $F_{\mathcal A}(x)$ correctly reflects the presence of an infinite barrier.

This issue is even more likely to arise in higher-dimensional settings, where such artifacts are also considerably more difficult to diagnose when relying solely on standard FES.
For example, consider a three-dimensional potential energy surface with coordinates $(x,y,z)$, projected onto the $(x,y)$-plane. In such a projection, two reaction channels may appear accessible, while in reality one of them is blocked by an impenetrable barrier oriented along the $z$-direction. The unconditional FES in the $(x,y)$-plane will then incorrectly suggest that both channels are viable. In contrast, the conditional FES correctly identifies that one of the channels is inaccessible.

These examples highlight that conditional free energies can more faithfully represent the underlying kinetics when projections onto reduced coordinates obscure key dynamical features. While such misrepresentations of standard FES can be mitigated by choosing appropriate collective variables, identifying these is nontrivial, and they may involve complex nonlinear functions of distances, angles, and other degrees of freedom, making low-dimensional FES representations difficult to interpret intuitively.
This motivates further investigation of these conditional free energies through a set of numerical examples.

\section{Results}

To illustrate history-dependent free energy, we performed $\infty$RETIS simulations on a Langevin particle in external potentials (Sects.~\ref{subsec:flat}–C) and on a fully atomistic membrane system using the infRETIS software~\cite{infretissoftware}, interfaced with ASE (Sect.~\ref{subsec:flat}), its internal MD engine~\cite{infretissoftware} (Sects.~\ref{Sec:dw}–C), or GROMACS~2024.3~\cite{GROMACS2024.3manual} (Sect.~\ref{sec:mem}). Reduced units ($k_B=1$, $m=1$) were used in Sects.~\ref{subsec:flat}–C unless stated otherwise. Langevin simulations comprised $2\times10^5$ MC steps, except for the $B\rightarrow A$ transition in Sect.~\ref{sec:2D}, which used $3\times10^5$ steps to sample the largest barrier. In Sect.~\ref{subsec:flat}, each MC step consisted of standard shooting with early rejection for all path ensembles. From ~\ref{Sec:dw} onward, high-acceptance WF moves with four subtrajectories were used, except $[0^+]$ and $[0^-]$, which again used standard shooting with early rejection~\cite{WF}. The subcycle parameter, controlling how often $\lambda$ is evaluated, was 1 for Langevin simulations and 100 for the membrane system. Time steps were $dt=0.002$ (\ref{subsec:flat}), 0.01 (\ref{Sec:dw}-C) in reduced units, and 2 fs (\ref{sec:mem}).
The systems discussed in \ref{subsec:flat}-C are publicly available on GitHub, together with Python scripts used to generate the results~\cite{infretis_infentory}. 

\subsection{Flat and cosine bump potentials}
\label{subsec:flat}
We study a Langevin particle evolving on either a flat potential or a cosine-shaped barrier, as also studied in Ref.~\citenum{ghysels2021exact} with a temperature $T=1$.
\begin{equation}\label{eq:u1d}
u_{\mathrm{cos}}(x)=
\begin{cases}
\tfrac{1}{2}V_0\left[\cos\!\left(\tfrac{\pi x}{a}\right)+1\right], & |x|\leq a,\\
0, & |x|>a.
\end{cases}
\end{equation}
The flat case corresponds to $V_0=0$, and the cosine barrier to $V_0=1$ ($1\,k_BT$) with $a=0.1$.
The order parameter \(\lambda=x\) is used with interfaces 
at $-0.1$, 0.0, and 0.1, and an additional 
$\lambda_{-1}$
interface
required for unbounded systems\cite{ghysels2021exact} 
was set at $-0.2$.

\begin{figure}[ht!]
    \centering
\includegraphics[width=\linewidth]{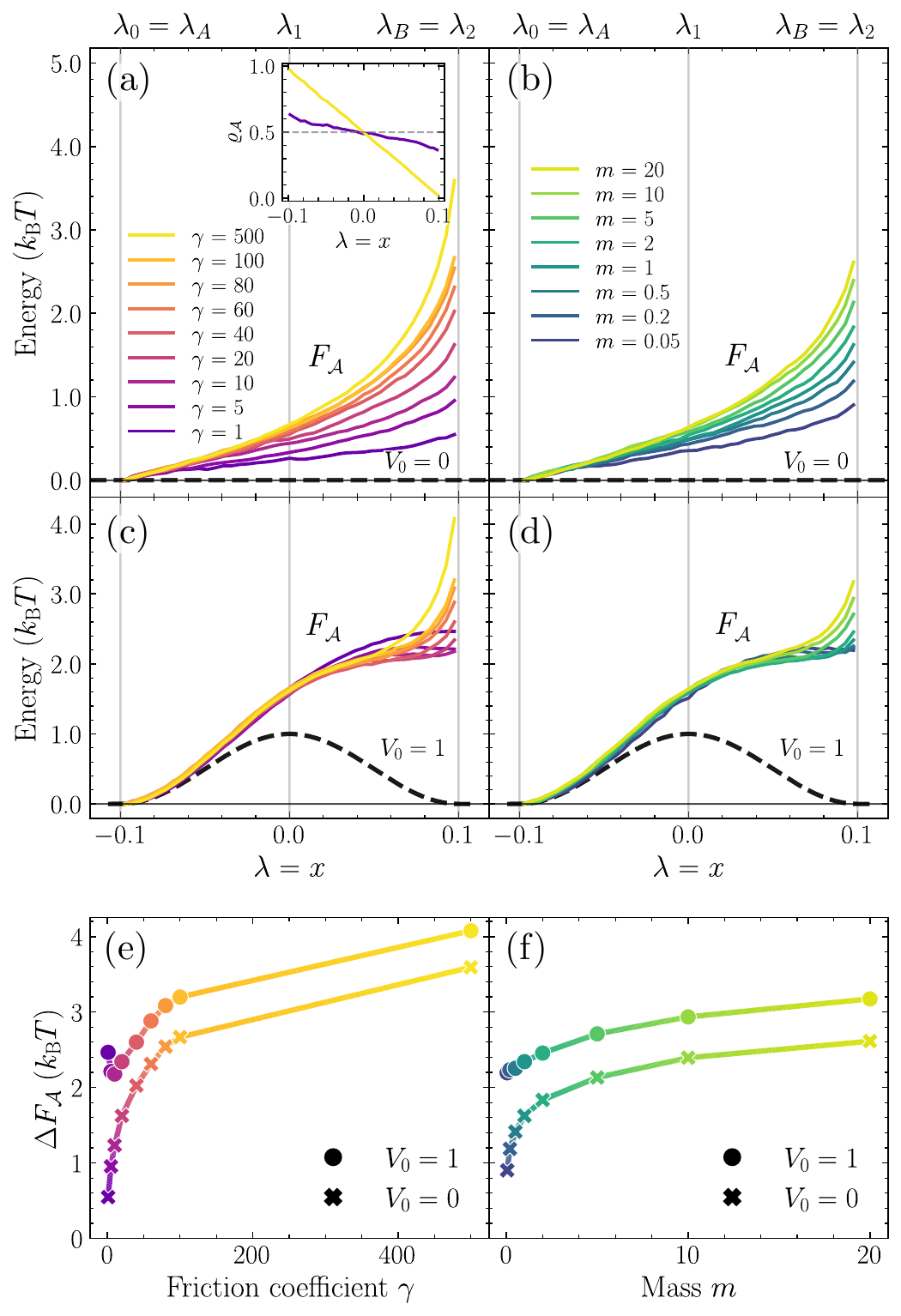}
    \caption{
    (a)--(d): Conditional free energy $F_\mathcal{A}(\lambda)$ for the flat potential (a,b) and the cosine bump  (c,d). The inset in (a) depicts $\varrho_\mathcal{A}(\lambda)$ for $\gamma=1$ and 500 for the flat potential.
    $u_\text{\rm cos}(x)$ is drawn with a black dashed line. 
    Panels (a) and (c) show results for various friction $\gamma$ values with fixed mass $m=1$, while panels (b) and (d) show results for various mass values $m$ with fixed $\gamma=20$.
    (e)--(f): $\Delta F_\mathcal{A}$ as a function of $\gamma$ and $m$.
    }
    \label{fig:1dgrid}
\end{figure}

Both potentials are simulated across a broad range of friction coefficients $\gamma$
(inverse time units)
with the mass kept constant at \(m=1\), and for different particle masses at constant friction \(\gamma=20\). 
Figs.~\ref{fig:1dgrid}a--d show 
\(F_\mathcal{A}(x)\) 
together with the potential $u_\text{cos}(x)$, which, for a one-dimensional system, coincides with the unconditional free energy.
Figs.~\ref{fig:1dgrid}e--f further examine the difference \(\Delta F_\mathcal{A} = F_\mathcal{A}(\lambda_B)-F_\mathcal{A}(\lambda_A)\)
as a function of \(\gamma\) and \(m\).

For the flat potential, the conditional  probability histogram $\varrho_{\mathcal A}(\lambda)$ (Eq.~\ref{eq:propA}, shown as the inset in Fig.~\ref{fig:1d_dw}a), scaled such that the constant total distribution $\varrho(\lambda)$ (Eq.~\ref{eq:scaling}) equals unity, decreases linearly from $1-\delta$ at the left to $\delta$ at the right, crossing $0.5$ at the center, with $\delta$ related to the 
crossing probability of reaching the right boundary when starting from the left.
Using Eq.~\ref{eq:FA}, the conditional free energy in Figs.~\ref{fig:1dgrid}a--b therefore takes the form 
$F_\mathcal{A}(\lambda)\approx - k_B T \ln\bigl((1-\delta-(1-2\delta) z)/(1-\delta)\bigr)$, with $z = (\lambda-\lambda_A)/(\lambda_B-\lambda_A)$.

Increasing the mass or friction decreases $\delta$ and thus raises $\Delta F_{\mathcal A}$ (Figs.~\ref{fig:1dgrid}e–f). This follows from the decrease of the diffusion coefficient, $D = k_B T/(m\gamma)$, with increasing $m$ or $\gamma$, which lowers the crossing probability.\cite{PPTIS}
The conditional free energy therefore reflects an effectively increased kinetic barrier. This kinetic barrier is physically meaningful, as it becomes harder to transfer from $\lambda_A$ to $\lambda_B$ when the diffusion coefficient is small, though this barrier remains invisible in the unconditional free energy.

When simulating a cosine-shaped potential (Figs.~\ref{fig:1dgrid}c--d), the initial rise of $F_\mathcal{A}(x)$ follows the barrier slope but crosses the top at $k_B T \ln 2$ above the underlying potential,
a universal feature of symmetric barriers reflecting that at the center, half the phase points belong to $\mathcal A$ and half to $\mathcal B$. Beyond this point, the curves initially follow the same increasing trend regardless of $m$ or $\gamma$, then rise steeply at the end, with a steeper slope for larger $m$ and $\gamma$.

In the low-$\gamma$ regime ($\gamma \le 10$), momentum dominates, allowing the particle to reach $\lambda_B$ unimpeded after crossing the barrier top. Here, $\Delta F_\mathcal{A}$ shows a slight decrease with increasing $\gamma$, as higher friction reduces momentum on the downslope, increasing phase-point density there and lowering $F_\mathcal{A}$. This leads to a minimum in $\Delta F_\mathcal{A}$ around $\gamma = 10$ (Fig.~\ref{fig:1dgrid}e).
At higher $\gamma$, diffusive effects dominate, and for $\gamma > 40$, $\Delta F_\mathcal{A}$ evolves similarly to the flat potential case. The trend with mass is analogous (Fig.~\ref{fig:1dgrid}f), except that the low-$\gamma$ dip in $\Delta F_\mathcal{A}$ is absent for small $m$, presumably because the constant friction $\gamma=20$ is already too large to allow this ballistic effect.

We also observed that the curvatures in Figs.~\ref{fig:1dgrid}c–d can be affected by the subcycle setting in $\infty$RETIS, which controls after how many MD steps $\lambda$ is evaluated along  trajectories ($\text{subcycle}=1$ is used here). The subcycle is known to influence the flux and the crossing probability $P_A(\lambda_B|\lambda_A)$, although these effects cancel in the rate calculation.\cite{Bolhuis2021} Nevertheless, $F_\mathcal{A}(x)$ shows some dependence on the subcycle near the absorbing boundary, where crossings of $\lambda_B$ may be missed for large subcycles.

\subsection{Double-well potential}
\label{Sec:dw}

\begin{figure}[htb]
    \centering
    \includegraphics[width=0.99\linewidth]{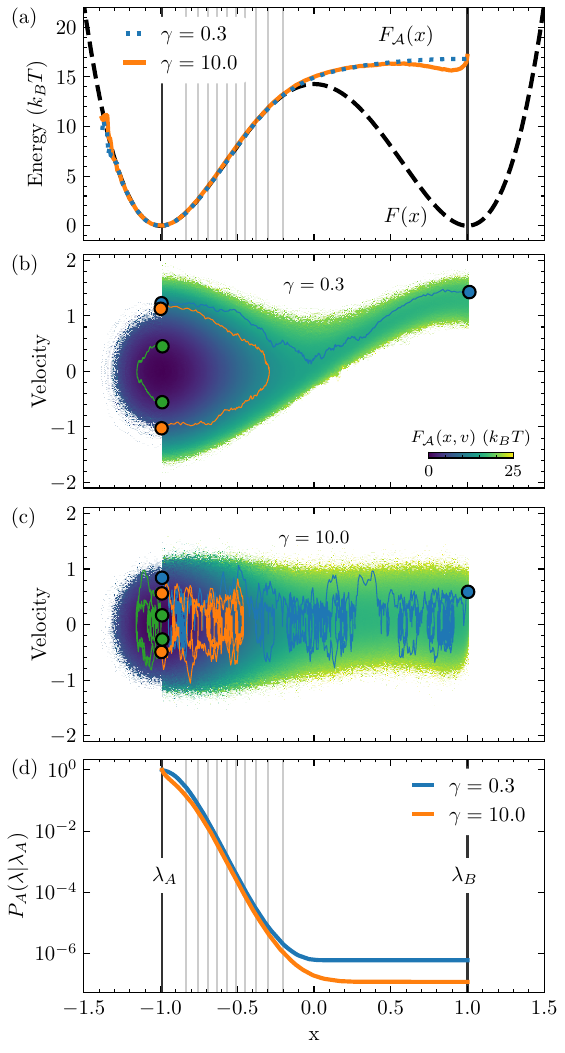}
    \caption{Double-well potential $u_{\rm dw}(x)$ of Eq.~\ref{eq:dw}. Reduced units. 
    (a): The unconditional free energy $F(x)$ (dashed black line) and the conditional free energy $F_\mathcal{A}(x)$ for $\gamma=0.3$ and $\gamma=10$.
    The utilized interfaces are shown in vertical gray/black in (a) and (d).
    (b) and (c): Heatmap of the $x$ and velocity conditional free energy $F_\mathcal{A}(x,v)$ for $\gamma=0.3$ and $\gamma=10$, with example reactive (blue), nonreactive (orange), and state A (green) trajectories also shown. (d): The 
    crossing probability $P_A(\lambda|\lambda_A)$ for the two $\gamma$ simulations.
    }
    \label{fig:1d_dw}
\end{figure}

A double-well potential $u_{\rm dw}$ is considered,~\cite{vanErp2012}
\begin{equation}
\label{eq:dw}
   u_{\rm dw}(x)=x^4-2 x^2
\end{equation}
at temperature $T=0.07$ for two friction coefficients, $\gamma=0.3$ and $\gamma=10$. We use $x$ as the order parameter and record the particle velocity as a secondary collective variable.
The utilized interfaces are shown in Figs.~\ref{fig:1d_dw}a, d.

Fig.~\ref{fig:1d_dw}a shows $F_\mathcal{A}$ for the two different $\gamma$ values as a function of $x$.  As in the previous case, the unconditional free energy curve coincides with the underlying 1D potential energy. To the left of the maximum, $F_\mathcal{A}(x)$ follows the unconditional free energy much more closely than in Fig.~\ref{fig:1dgrid}c–d. This is due to a much higher barrier: while both barriers equal 1 in reduced units, the effective barrier in thermal units is $1/0.07 \approx 14.29~k_B T$, compared with $1~k_B T$ in the previous case.
As before, the conditional free energy is shifted upward by $k_B T\ln 2$ relative to the unconditional free energy at the transition state $x=0$, and increases slightly further beyond it. For $x>0$, this increase is nearly identical for both friction values, despite their fundamentally different dynamics, as clarified below.

Figs.~\ref{fig:1d_dw}b–c show the two-dimensional conditional free energy as a function of $x$ and velocity $v$, together with representative paths. In the low-friction case (Fig.~\ref{fig:1d_dw}b), a pronounced asymmetry 
with respect to the $v=0$ axis
 appears for $x>0$, reflecting ballistic motion: once the barrier is crossed, trajectories accelerate downhill and no longer attain negative velocities. This agrees with the crossing probability in Fig.~\ref{fig:1d_dw}d reaching a flat plateau almost immediately after the transition state.

In contrast, for $\gamma=10$ the plateau in the crossing probability is shifted to larger $x$, indicating that many trajectories still return to state $A$ after crossing $x=0$. Consistently, the conditional free energy surface in the high-friction regime (Fig.~\ref{fig:1d_dw}c) remains nearly symmetric along the velocity axis.
It is therefore striking that the conditional free energy curves for $\gamma=0.3$ and $\gamma=10$ are nearly identical up to $x\approx0.4$, even though the underlying mechanisms differ for the increase in 
$F_\mathcal{A}(x)$ for $x>0$. At high friction, the increase reflects continued recrossings that reduce the number of surviving  paths as they return $A$ to instead of reaching further.  At low friction,
however, nearly all  trajectories  crossing $x=0$  move irreversibly toward $B$, but rapid acceleration lowers the phase-space density sampled by the histogram. Although opposite in nature, these two effects produce nearly the same free-energy profile in the region $0 < x < 0.4$.

It is therefore clear that an increase in the conditional free energy is not always a signature of a barrier. While this interpretation is valid for the $\gamma = 10$ case, where it reflects that many trajectories fail to reach the final state, 
in the low-friction regime, the increase does not indicate hindered progress but rather the reverse. This implies that conclusions based solely on the conditional free energy must be drawn with care in the low-friction regime; however, by jointly analyzing complementary observables, such as the crossing probability and free-energy surfaces that include velocity along the configurational collective variable, the full dynamical picture becomes apparent, allowing physically meaningful conclusions to be drawn.

Fig.~\ref{fig:1d_dw}a shows a slight deviation in $F_\mathcal{A}(x)$ for $x>0.4$, where the high-friction curve displays a small initial decline followed by a steep rise near the endpoint, while the low-friction curve approaches a plateau. At this stage, both systems have passed the point of no return, as indicated by the flat crossing probabilities in Fig.~\ref{fig:1d_dw}d. Nevertheless, for $\gamma=10$ the transition paths still exhibit pronounced back-and-forth motion along $x$ before reaching state $B$ (Fig.~\ref{fig:1d_dw}c). The initial decline reflects dissipation of kinetic energy gained on the downhill slope, whereas the final steep rise results from the absorbing boundary at $\lambda_B$. Because high-friction trajectories undergo multiple reversals, histogram bins are revisited by the same path, but once $\lambda_B$ is crossed trajectories are terminated, preventing multiple counts in bins close to $\lambda_B$. This provides another example in which an increase in $F_\mathcal{A}(x)$ does not correspond to the presence of a barrier.

Figs.~\ref{fig:1d_dw}b–c also show a visual discontinuity at $\lambda_A$, most pronounced in the low-friction case, where results from the $[0^-]$ and $[i^+]$ ensembles are joined. A small physical discontinuity may arise for low barriers since a tiny fraction of phase points just to the right of $\lambda_A$ belongs to $\mathcal{B}$, whereas this fraction is zero by definition to the left. Such an effect is not noticeable for crossing probabilities as small as $10^{-6}$. The observed discontinuity instead reflects missing sampling: the $[0^-]$ ensemble rarely generates trajectories with extreme kinetic energies, while the corresponding high-velocity contributions on the right of $\lambda_A$ are recovered only through reweighting with $[i^+]$ ensembles. This sampling gap also explains why splitting methods such as FFS\cite{Allen2005} fail to converge for this system. These methods rely on forward propagation from MD-generated points, but high-velocity configurations at $\lambda_A$ are too rare to be sampled directly, even though they dominate successful transfer, leading to underestimated rates and an incorrect mechanistic picture.\cite{vanErp2012}
To capture these rare but essential contributions and obtain accurate results, FFS requires a velocity-dependent order parameter.\cite{contourFFS}

\subsection{2D potential}
\label{sec:2D}
\begin{figure*}[htb]
    \centering
    \includegraphics[width=0.97\linewidth]{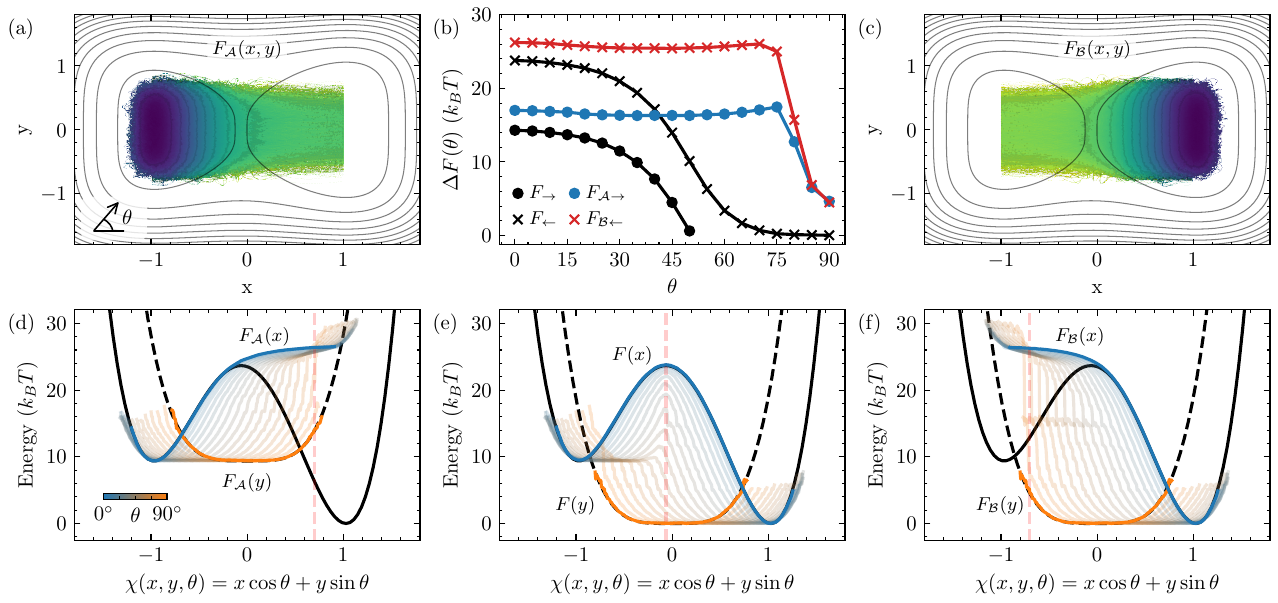}
    \caption{The 2D potential $u_\text{2D}$ (Eq.~\ref{eq:2dangle}) is shown by the black contour lines in sub-panels (a) and (c). 
    The 1D $x$ and $y$ projections are shown by the black solid ($F(x)$) and dashed ($F(y)$) lines in (d), (e) and (f). 
    The computed 2D conditional free energies $F_\mathcal{A}(x,y)$ and $F_\mathcal{B}(x,y)$ are shown in (a) and (c) as a heatmap (with maximum values $\approx24$ and 34 $k_BT$).
    The 1D conditional free energy profiles $F_\mathcal{A}(x)$ (blue) and $F_\mathcal{A}(y)$ (orange) for the forward transition are shown as lines in (d), 
    and similarly for $F_\mathcal{B}(x)$ (blue) and $F_\mathcal{B}(y)$ (orange) for the backward transition in (f). 
    Other colored lines (blue shade to orange shade) in (d)-(e)-(f) are $F_\mathcal{A}(\chi;\theta)$, $F(\chi;\theta)$, and $F_\mathcal{B}(\chi;\theta)$ for a series of parameter values $\theta$, i.e.\ for changing definitions of the $\chi$-axis.
    The conditional zero free energy and the unconditional $y$ projection in (d) is shifted to the local left minima in (d). 
    The vertical red dashed lines in (d), (e) and (f) 
    serve as reference points for determining the barrier heights $\Delta F(\theta)$ shown in (b).
    }
    \label{fig:angle_combo}
\end{figure*}

To investigate the relationship between the order parameter and the resulting conditional free energy barrier, we  consider the same $\gamma=0.3$ Langevin dynamics 
with stable state boundaries $\lambda_A$ and $\lambda_B$
as in Sect.~\ref{Sec:dw}, but with the double well replaced by a two-dimensional potential  described by Dietschreit et al.\cite{Dietschreit2022}
\begin{equation}
\label{eq:2dangle}
u_{\rm 2D}(x, y) = 1.3128 (y^4 + x^4 - 2x^2 - 0.25x)
\end{equation}
where the prefactor is chosen to match the energy difference between the left minimum and the transition state in the double-well example.
Illustrated by the black contour lines in Fig.~\ref{fig:angle_combo}a (or \ref{fig:angle_combo}c), the ideal order parameter describing the pathway between the two metastable states is simply the $x$ coordinate of the particle, where $y$ is an orthogonal coordinate that does not contribute to the reaction progress.
This is also observed in the one-dimensional $x$ and $y$ unconditional free energy
shown in Figs.~\ref{fig:angle_combo}d-f,
where the black line displaying asymmetric double well is the $x$ projection $F(x)$, and the black dashed line displaying the single well is the $y$ projection $F(y)$.

We perform forward and backward $\infty$RETIS simulations with $x$ and $-x$ as the order parameters, respectively.
The conditional free energy in the forward and backward simulation, $F_\mathcal{A}(x)$ and $F_\mathcal{B}(x)$, are shown in Figs.~\ref{fig:angle_combo}d and \ref{fig:angle_combo}f (blue line). They exhibit the same trend as the double-well conditional free energy $F_\mathcal{A}(x)$ in Fig.~\ref{fig:1d_dw}a.

As in Dietschreit et al.,\cite{Dietschreit2022} we investigate how the conditional and unconditional free energy changes when projected onto a collective variable $\chi$ describing a rotating axis from $x$ to $y$,
\begin{equation}
    \chi(x,y,\theta)= x\cos\theta + y\sin\theta
\end{equation}
where the $\chi$ definition depends on a parameter $\theta$, with $\chi(x, y,\theta)=x$ if $\theta=0^\circ$ and $\chi(x, y,\theta)=y$ if $\theta=90^\circ$. 
The angle $\theta$ definition is also illustrated in Fig.~\ref{fig:angle_combo}a, where the arrow represents the orientation of the $\chi$-axis.
The projection on the $\chi$-axis is now used at the CV in Eq.~\ref{eq:FA}.
Figs.~\ref{fig:angle_combo}d and \ref{fig:angle_combo}f show the the conditional free energy $F_\mathcal{A}(\chi;\theta)$ and $F_\mathcal{B}(\chi;\theta)$, respectively, for a series of angles $\theta$.
As the angle increases (the curves turn more orange), the projected conditional free energy $F_\mathcal{A}(\chi;\theta)$ appears to retain much of its initial $\theta=0^\circ$ character.
Primarily only as the angle reaches $\theta=90^\circ$ does the projected conditional $F_\mathcal{A}(y)$ and $F_\mathcal{B}(y)$ free energies collapse to the $y$ projected unconditional free energy $F(y)$ (orange line). 
In comparison, the unconditional free energy $F(\chi;\theta)$ shown in Fig.~\ref{fig:angle_combo}e appears to smoothly transition from the $x$ to the $y$ projection with increasing $\theta$.

Fig.~\ref{fig:angle_combo}b shows the barrier heights $\Delta F(\theta)$, defined as the difference in energy at 
the barrier and the minimum in state $A$ or $B$ for the $A\rightarrow B$ and $B\rightarrow A$ reactions, respectively, of the conditional and unconditional free energy curves. While the unconditional free energy exhibits a clear barrier located at the local maximum (red dashed line in Fig.~\ref{fig:angle_combo}e), the conditional curves form a plateau rather than a distinct maximum (Fig.~\ref{fig:angle_combo}d,f). The values $\chi=0.7$ and $\chi=-0.7$ were therefore chosen to measure the plateau height, as indicated by the vertical red dashed lines in Figs.~\ref{fig:angle_combo}d,f. 

As shown in Fig.~\ref{fig:angle_combo}b, the free energy barriers of the conditional curves remain roughly constant up to $75^\circ$ and do not decline like the standard unconditional free energy curve. The latter behavior reflects the well-known effect of a sub-optimal reaction coordinate, where free energy barriers are underestimated relative to the true physical barriers. In such cases, methods like umbrella sampling (US)\cite{US} may miss relevant configurations near the transition state, leading to hysteresis or incomplete sampling. In contrast, path sampling maintains trajectories connected to the stable state of origin, effectively sampling along the conditional free energy surfaces, $F_{\mathcal A}$ or $F_{\mathcal B}$, and capturing the true transition behavior. This is one of the reasons postulated in Ref.~\citenum{vanErp2006} why TIS rate calculations are more efficient than standard free energy–based reactive flux methods if the reaction coordinate is poor.

\subsection{Membrane permeation of drug molecule}
\label{sec:mem} 

To demonstrate (un)conditional free energies in a biologically relevant system, we analyzed permeation trajectories of 5-aminolevulinic acid (5-ALA) through a DPPC bilayer (64 lipids), taken from our earlier all-atom MD and path sampling study using the CHARMM36 lipid force field and TIP3P water model under NPT conditions at 323 K and 1 bar~\cite{safaei2025exact}. The reaction coordinate $\lambda$ was defined as the displacement along the membrane normal ($z$-axis) between the centers of mass of 5-ALA and the membrane. Thirty interfaces $\lambda_i$ were placed between $\lambda_0=\lambda_A=-24~\text{\AA}$ and $\lambda_B=24~\text{\AA}$, defining 30 ensembles in total (see Ref.~\citenum{safaei2025exact} for details). An additional interface at $\lambda_{-1}=-34~\text{\AA}$ was introduced to prevent periodic boundary crossings, as discussed in Appendix~\ref{sec:cor_endR}.

\begin{figure}[htb!]
    \includegraphics[width=1\linewidth]{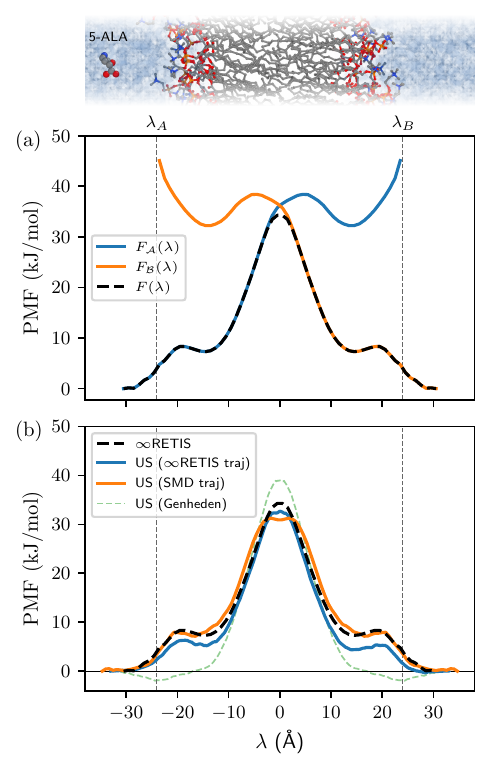}
    \caption{a) (Un)conditional free energies of the drug molecule (5-ALA) permeating through the DPPC membrane as a function of the order parameter ($\lambda$). b) Comparison of the unconditional free energies obtained from $\infty$RETIS and umbrella sampling simulations starting from different initial trajectories. All free energy profiles are shifted to zero in the water phase. The PMF profiles obtained from the SMD trajectory and from Genheden and Eriksson~\cite{genheden2016estimation} were symmetrized to span the full transition from state $A$ to state $B$.}   
    \label{Fig:5ALA_1D_FE}
    \end{figure}

In total, $13,635$ MC cycles were performed, of which $11,878$ individual trajectories were accepted. The first $970$ trajectories were treated as path sampling equilibration and excluded from further analysis. Phase points from the remaining $10,908$ accepted trajectories were collected with their proper weights to compute the conditional free energy $F_\mathcal{A}(\lambda)$, shown in Fig.~\ref{Fig:5ALA_1D_FE}a. Since membrane permeation is symmetric for this bilayer, the conditional free energy $F_\mathcal{B}(\lambda)$ was obtained by mirroring $F_\mathcal{A}(\lambda)$ about $\lambda=0$. From these profiles, the unconditional free energy $F(\lambda)$ was calculated using Eq.~\ref{eq:free} (black dashed line in Fig.~\ref{Fig:5ALA_1D_FE}a). The unbounded-system correction factor $\xi$ was applied according to 
Appendix~\ref{sec:cor_endR}.

The unconditional free energy reflects the behavior of an ideal, perfectly symmetric membrane. In practice, permeation induces local membrane disturbances that can lead to hysteresis, making the conditional free energy possibly a more reliable estimate of the effective barrier. This interpretation is supported by Ref.~\citenum{safaei2025exact}, which shows that although trajectories reach the membrane center, only about 20\% proceed to state~$B$. This explains the continued increase of $F_\mathcal{A}(\lambda)$ beyond $\lambda=0$, similar to the behavior discussed in Sec.~\ref{subsec:flat}-C. Here, it is followed by a pronounced dip corresponding to a short-lived metastable state around $14.5$~\AA, and a steep rise near the absorbing boundary, consistent with the other simulations.

The unconditional free energy $F(\lambda)$, or potential of mean force (PMF), is compared with US simulations using 136 umbrella windows spaced at 0.5 Å. A harmonic bias of 1000 kJ mol$^{-1}$ nm$^{-2}$ maintained the drug at the target reaction coordinate. Each window was simulated for 20 ns, discarding the first 10 ns as equilibration. The PMF was constructed using the GROMACS WHAM implementation\cite{hub2010g_wham} with a convergence tolerance of $1\times10^{-6}$ kJ/mol. Two different approaches were used to generate initial configurations for the US simulations.

 In one set, the initial trajectory came from a steered MD (SMD) simulation where 5-ALA was pulled at 0.1 $\text{\AA}$/ns from bulk water ($\lambda=-34$ Å) to the membrane center ($\lambda=0$ Å). The molecule was then restrained at the center with a 5000 kJ/mol/nm$^2$ force for 50 ns to equilibrate the membrane and relax lipid distortions. After release, the trajectory exiting the membrane naturally served as the US initial path. As this trajectory covers only the path from the center to bulk water, the resulting PMF was symmetrized.

In the second set of US simulations, a transitioning trajectory from the $\infty$RETIS simulation, taken from the highest path ensemble $[28^+]$ after ~3400 accepted paths, was used as the initial trajectory. Unlike the first US set, this RETIS trajectory already spans $A\to B$, so the resulting free energy did not require symmetrization.

In Fig.~\ref{Fig:5ALA_1D_FE}b, the PMF profiles are also compared with a previously published profile by Genheden and Eriksson\cite{genheden2016estimation} for the same system. Since that study employed a different force field (GAFF), the profiles are not quantitatively comparable and are included only to provide qualitative context for the overall shape of the free energy landscape.

When comparing the overall shape of the PMF profiles in Fig.~\ref{Fig:5ALA_1D_FE}b, the US profile obtained from the SMD trajectory is very similar to the $\infty$RETIS free energy profile, except near the center of the membrane. This difference likely occurs because the initial SMD trajectory covers only the region from the membrane center to the bulk water, which leads to incomplete sampling of the full permeation pathway. Such partial PMF profiles are a common limitation of nonequilibrium methods like SMD, where pulling a molecule through the entire membrane can cause noticeable structural distortions and symmetrization of the partial profile often hides poor convergence. \cite{Paloncov2012,Pokhrel2018,Sousa2023}

This limitation is mitigated in the US simulations initialized from the $\infty$RETIS trajectory, which represents a continuous transition from side $A$ to side $B$ across the membrane. While the overall agreement between the $\infty$RETIS profile and the US($\infty$RETIS) profile is good, differences are observed at specific points along the reaction coordinate. In particular, at the metastable states around $\pm14.5$~Å, the free energies are 7.3~kJ/mol for $\infty$RETIS and 5.2~kJ/mol for US($\infty$RETIS), while at the top of the barrier they are 34.3~kJ/mol and 32.7~kJ/mol, respectively. These differences highlight the sensitivity of PMF estimates to the choice of initial pathway and sampling strategy, in line with more general limitations of US discussed in the literature\cite{aho2024all}, yet we cannot say undoubtedly which free energy curve is the most accurate due to the lack of knowledge of the exact profile.

\section{Conclusion}

$\infty$RETIS  is a highly parallelizable method that boosts path sampling efficiency by combining asynchronous replica exchange in the infinite-swap limit with advanced shooting moves that decorrelate rapidly and maintain high acceptance rates, while sampling a  biased path distribution. In this framework, the focus shifts from ensembles containing separate sets of paths to the paths themselves, each represented by a frequency list indicating its fractional visits to different ensembles.

This output requires a generalized treatment of path reweighting, which forms the focus of the first part of this article. We derive a generic relation assigning each path an overall WHAM weight (Eqs.~\ref{eq:barLw} and~\ref{eq:Lamdef}), combining ensemble-specific estimates while correcting for the biased distribution, fractional contributions, and ensemble overlap, thereby minimizing statistical error. 
These WHAM weights are obtained through a single fast forward iteration, without the self-consistent iterative procedure required in most  WHAM approaches.
Consequently, path observables can be computed as weighted averages over all $\infty$RETIS paths without reference to individual ensembles (Eq.~\ref{eq:barLamLam}).

Phase-space and configurational-space ensemble averages follow directly from path averages by collecting time slices along the paths, yielding weighted averages over phase points without bookkeeping of their path or ensemble. These points belong to the overall state $\mathcal{A}$, which includes the stable state $A$ as well as barrier configurations more recently in $A$ than in $B$. Averages from a single $\infty$RETIS simulation thus correspond to history-dependent conditional averages. If the reverse reaction is also studied, results can be combined to recover unconditional averages. 
For symmetric barriers, this extra simulation is unnecessary, and a single $\infty$RETIS simulation provides both conditional and unconditional results.

In the second part of this article, we apply the reweighting schemes to compute and analyze free energy profiles, including both standard (unconditional) profiles from forward and backward $\infty$RETIS simulations and conditional profiles, for a set of low-dimensional Langevin systems as well as a full-atom model of membrane permeation.  

While unconditional free energies provide the conventional thermodynamic picture, conditional profiles contain additional information not visible in standard curves. 
Due to their history dependence, conditional free energies reflect kinetic effects and therefore change when parameters such as particle mass or friction in Langevin dynamics are altered, modifications that leave unconditional free energies unchanged. Conditional profiles can reveal kinetic barriers caused by diffusive dynamics, where trajectories fail to reach the product state even without a true configurational or entropic barrier. 
Yet, caution is required  as not every increase in conditional free energy reflects a physical barrier; for example, downhill acceleration in the low-friction limit or absorbing boundaries in the high-friction limit can also produce increases.

A key feature of conditional barriers is their insensitivity to the choice of collective variables. Since these barriers are typically within $2$–$3 k_B T$ of the unconditional barriers corresponding to the true reaction coordinate, they provide meaningful estimates even when hysteresis leads to severe underestimation in standard profiles. Moreover, because both conditional and unconditional free energies from $\infty$RETIS can be projected onto any collective variable, not just the progress coordinate $\lambda$ used to define the interfaces, they can be employed in a variational approach to identify the correct reaction coordinate and corresponding unconditional barrier. While this is in principle also possible with biasing methods such as umbrella sampling, path sampling ensures sufficient sampling near the transition state, which is often lacking in US simulations based on suboptimal coordinates.

Another further methodological development is that path data generated using different choices of the progress coordinate $\lambda$ can be consistently combined through multi-set path reweighting.\cite{mbar_pr} This also provides flexibility with respect to the choice of order parameter used to place interfaces, and it can be effectively combined with adaptive data-driven approaches for improving collective variable or interface definitions.\cite{jcp_ml1,jcp_ml2,adaptive_cv}
To conclude, we derived a generalization of path reweighting that works for $\infty$RETIS path data and demonstrated how it can be used to compute conditional free energy profiles that capture history-dependent effects invisible in standard profiles. These profiles reveal subtle features of the barrier region and the reaction dynamics. Moreover, they provide quantitative results that are robust with respect to the quality of the reaction coordinate, thereby overcoming a key limitation of traditional free energy methods.

\appendix

\section{Unbounded systems with $\lambda_{-1}$ interface}
\label{sec:cor_endR}
In Eq.~\ref{eq:phase_av_overall}, it was assumed that a hypothetical infinitely long MD trajectory generates an equal number of valid trajectory segments in the $[0^-]$ and $[0^+]$ ensembles, as defined by Eq.~\ref{Eq:cond}.
This assumption is valid for bounded systems, where the free energy rises steeply
for order-parameter values smaller than the minimum of the stable state $A$
(e.g.\ in the region $\lambda<-1.3$ in Fig.~\ref{fig:1d_dw}A). 
A typical example is a chemical reaction with a bond-length order parameter, which is bounded both physically, due to the steep increase in potential energy upon compression, and mathematically, since the bond length cannot be negative.
In contrast, for membrane permeation, where the position of a permeant relative to the membrane is used as the reaction coordinate, no such left bound exists, and the separation can formally extend to minus infinity.
In Ref.~\onlinecite{ghysels2021exact}, this issue was addressed by introducing an additional interface $\lambda_{-1}$ to the left of $\lambda_0$ and redefining the $[0^-]$ ensemble to include paths that may start and end at either interface. With corresponding modifications to the RETIS expressions, this enables an exact computation of the permeation constant.

Since the endpoint of a $[0^-]$ path is no longer necessarily the starting point of a $[0^+]$ path, an infinitely long MD trajectory will generally produce more segments in the modified $[0^-]$ ensemble. This requires adapting Eq.~\ref{Eq:cond} using the fraction of $[0^-]$ paths that end at $\lambda_0$,
\begin{align}
\xi = \left\langle \mathbbm{1}_{\rightarrow}(X) \right\rangle_{\bar{0}},
\end{align}
where the indicator function equals 1 (and 0 otherwise) if the path $X$ ends at $\lambda_0$, irrespective of its starting interface.
The phase space average in ${\mathcal A}$, Eq.~\ref{eq:phase_av_overall}, is then adapted to
\begin{align} 
    \left\langle o(x) \right\rangle_{\mathcal A}
    &=
    \frac{
        \left\langle O(X; o) \right\rangle_{\bar{0}}
        + \xi
        \left\langle O(X; o) \right\rangle_{0}
    }{
        \left\langle \hat{L}(X) \right\rangle_{\bar{0}}
        + \xi
        \left\langle \hat{L}(X) \right\rangle_{0}
    } ,
    \label{eq:phase_av_overall_unbounded}
\end{align}
which in practice is achieved via Eq.~\ref{eq:ostar}, but with adapted phase space 
weights:  
\begin{align}
\Lambda^*(x) =
\begin{cases}
\bar{\Lambda}_j, &
\text{if } x \in \hat{X}_j \textrm{ and }  X_j \in [0^-], \\ 
\xi \Lambda_j, & \text{if } x \in \hat{X}_j \textrm{ and } X_j \notin [0^-], 
\end{cases}
\end{align}

\section{Derivation of WHAM weights $\Lambda_j$, Eq.~\ref{eq:Lamdef}}
\label{AppendixWHAM}
The aim is to derive $\Lambda_j$ of Eq.~\ref{eq:Lamdef} which are the WHAM weights of $\langle O\rangle_0$ in Eq.~\ref{eq:barLamLam}.
Let $X$ be a path that is valid for $[0^+]$ and
let 
$H_i(X)$ be the following indicator function:
\begin{align}
H_i(X)=
\begin{cases}
1 & \textrm{if } \lambda_i < \lambda_{\rm max}(X) \leq \lambda_{i+1}
\textrm{ for } i<n-1\\
  &
  \text{or }   \lambda_{\rm max}
  (X) >\lambda_{n-1}  \textrm{ for } i=n-1\\
0 & \text{otherwise }.
\end{cases} \label{eq:Hi}
\end{align}
where $[(n-1)^+]$ denotes the ensemble with the highest index employed in RETIS and $\infty$RETIS. 
The indicator function $H_i(X)$ equals 1 (and 0 otherwise) if the maximum $\lambda$ value along $X$ lies between $\lambda_i$ and $\lambda_{i+1}$, except for $i = n-1$, where it equals 1 whenever the
maximum exceeds $\lambda_{n-1}$, without any upper bound.

Using this definition, we can write
\begin{align}
\left \langle O(X) \right \rangle_0
=
  \left \langle O(X)
 \sum_{i=0}^{n-1} H_i(X) \right \rangle_0
=
\sum_{i=0}^{n-1}  \left \langle O H_i\right \rangle_0 \label{eq:sumHi}
\end{align}
The $i$th term in the sum can be computed from any ensemble $[k^+]$ with $k \leq i$, since in this range $\rho_0(X) H_i(X) = \rho_k(X) H_i(X)$. Indeed, the only distinction between the (non-normalized) distributions $\rho_0(X)$ and $\rho_k(X)$ lies in the minimal progress condition: $\rho_0(X)$ is nonzero if $\lambda_{\rm max}(X) > \lambda_0$ (Eq.~\ref{eq:1E}), while $\rho_k(X)$ requires $\lambda_{\rm max}(X) > \lambda_k$. Therefore, $\rho_0(X)$ and $\rho_k(X)$ only differ when $\lambda_0 < \lambda_{\rm max}(X) < \lambda_k$, but since $H_i(X) = 0$ in this case ($k \leq i$), both sides vanish, and hence the equality $\rho_0(X) H_i(X) = \rho_k(X) H_i(X)$ holds. Consequently, we can express $\left \langle O H_i\right \rangle_0$ as:
\begin{align}
  \left \langle O H_i\right \rangle_0 &=
  \frac{\int \rho_0(X) O(X) H_i(X) {\mathcal D}X}{
\int \rho_0(X)  {\mathcal D}X
  } \nonumber \\
  &=
  \frac{\int \rho_k(X) O(X) H_i(X) {\mathcal D}X}{
\int \rho_0(X)  {\mathcal D}X
  } \nonumber \\
  &=
  \frac{\int \rho_k(X) O(X) H_i(X) {\mathcal D}X}{
\int \rho_k(X)  {\mathcal D}X
  } 
   \frac{\int  \rho_k(X) {\mathcal D}X}{
\int \rho_0(X)  {\mathcal D}X
  } \nonumber \\
  &=
\left \langle O H_i\right \rangle_k P_A(\lambda_k|\lambda_0) \label{eq:A3}
\end{align}
when $k\le i$.
This demonstrates that the ensemble average $\left \langle O H_i\right \rangle_0$ can, in principle, be computed from the ensembles
$[0^+]$, $[1^+]$, up to $[i^+]$.
To combine the data of multiple ensembles, the weighted histogram analysis method (WHAM)\cite{WHAM1,WHAM2, WHAM3} can be used, which aims to get the best overall statistical accuracy by using a sensible weighted average:
\begin{align}
  \left \langle O H_i\right \rangle_0 
  &=
  \frac{ 
  \sum_{k=0}^i \alpha_k
\left \langle O H_i\right \rangle_k P_A(\lambda_k|\lambda_0)
}{
\sum_{k=0}^i \alpha_k
} \label{eq:WHAMweights}
\end{align}
where the weights $\alpha_k$ should be  selected
to approximately minimize the overall error. Ref.~\citenum{vanErp2016}  demonstrates that, under the  standard
WHAM assumptions, the optimal weights  can be expressed as
$\alpha_k = n_k \times [P_A(\lambda_k | \lambda_0)]^{-1}$, where $n_k$  represents
the number of trajectories sampled in $[k^+]$.
For $\infty$RETIS, $n_k$ 
is replaced by $\eta_k$, defined in Eq.~\ref{eq:etak},
such that Eq.~\ref{eq:WHAMweights} is transformed to:
\begin{align}
  \left \langle O H_i\right \rangle_0 
  &=
  \frac{ 
  \sum_{k=0}^i \eta_k
\left \langle O H_i\right \rangle_k 
}{
\sum_{k=0}^i \eta_k [P_A(\lambda_k|\lambda_0)]^{-1}
} \label{eq:WHAMweights2}
\end{align}
By substituting the $k$th
ensemble
result $\left \langle O H_i\right \rangle_k$ with Eq.~\ref{Eq:av2}, this becomes:
\begin{align}
  \left \langle O H_i\right \rangle_0 
  &=
  \frac{ 
  \sum_{k=0}^i \eta_k
\frac{\sum_{ j}  \mu_{jk} O(X_{j}) 
H_i(X_{j})/w_{k}(X_j)
 }{
\sum_{ j}  \mu_{jk}/w_{k}(X_j)
 }
}{
\sum_{k=0}^i \eta_k [P_A(\lambda_k|\lambda_0)]^{-1}
} \label{eq:tripsum}
\end{align}
Using the definitions of $Q_i$ and $t_{jk}$ in Eqs.~\ref{eq:Q} and \ref{eq:tjk} this leads to:
\begin{align}
  \left \langle O H_i\right \rangle_0 
  &=Q_i
  \sum_{k=0}^i 
\sum_{ j}  t_{jk} O(X_{j}) 
H_i(X_{j})
 \label{eq:OH_runest}
\end{align}

We will now return to the ensemble average $\left \langle O \right \rangle_0$ of Eq.~\ref{eq:sumHi}, in which the $\left \langle O H_i\right \rangle_0$ need to be substituted.
Comparison of Eqs.~\ref{eq:Hi} and \ref{eq:K} reveals the following relation between 
$K(X)$ and $H_i(X)$:
\begin{align}
H_i(X)= 
\delta_{i,K(X)}
\label{eq:HK}
\end{align}
where $\delta_{i,j}$ is the Kronecker delta.
This allows us to rewrite Eq.~\ref{eq:sumHi} as:
\begin{align}
  \left \langle O \right \rangle_0  
  &= \sum_{i=0}^{n-1} Q_i 
  \sum_{k=0}^{i}
  \sum_{ j} 
   t_{jk} O(X_{j})
  \delta_{i,K_j}
  \nonumber \\
&= \sum_{i=0}^{n-1} Q_i 
  \sum_{k=0}^{n-1}
  \sum_{ j} 
  t_{jk} O(X_{j})
  \delta_{i,K_j}
  \nonumber \\
  &= \sum_{ j} 
  \sum_{k=0}^{n-1}
  \sum_{i=0}^{n-1} Q_i 
  t_{jk} O(X_{j})
  \delta_{i,K_j}
  \nonumber \\
&= \sum_{ j} 
  \sum_{k=0}^{n-1}
  Q_{K_j}
  t_{jk} O(X_{j}) 
 \label{eq:extbound}
\end{align}
In the second line, we extended the upper limit of the summation over $k$ from $i$ to $n-1$. This is justified by the fact that $t_{jk}$ can only be nonzero if $\lambda_{\rm max}(X_j) > \lambda_k$, 
i.e.\  $K_j \ge k$.
Therefore, in the terms with $k>i$, $t_{jk}$ can be nonzero if $\lambda_{\rm max}(X_j) > \lambda_{i+1}$, i.e.\ when $K_j > i$. However, meanwhile $\delta_{i,K_j}$ equals zero for these terms,
and thus $t_{jk} \delta_{i,K_j}$ vanishes for all $k > i$, validating the extension of the summation limit.
This adjustment is crucial, as it allows us to change the order of the summations over 
$k$ and $i$ in the third line. 
In the fourth line,
the summation over $i$ vanishes, as only $i=K_j$ is non-zero due to the Kronecker delta.

Comparing Eqs.~\ref{eq:extbound} and \ref{eq:barLamLam}, we see these are equivalent if $\Lambda_j$ is defined as in Eq.~\ref{eq:Lamdef}, thus validating Eq.~\ref{eq:Lamdef}.

\section{Derivation of $P_A(\lambda_i|\lambda_0)$ in Eq.~\ref{eq:pp}} \label{app:B}

For $0 \leq k \leq i$, using $\rho_0(X)\,\theta(\lambda_{\rm max}(X)-\lambda_i) = \rho_k(X)\,\theta(\lambda_{\rm max}(X)-\lambda_i)$, we can write $P_A(\lambda_i | \lambda_0)$ analogously to Eq.~\ref{eq:A3}:
\begin{align}
     P_A(\lambda_i | \lambda_0) &= \left \langle 
    \theta(\lambda_{\rm max}(X) - \lambda_i)
    \right \rangle_0 \nonumber \\
  &=
  \frac{\int \rho_0(X) \theta(\lambda_{\rm max}(X) - \lambda_i) \, {\mathcal D}X}{
\int \rho_0(X) \, {\mathcal D}X
  } \nonumber \\
  &=
  \frac{\int \rho_k(X) 
   \theta(\lambda_{\rm max}(X) - \lambda_i)
  \, {\mathcal D}X}{
\int \rho_0(X) \, {\mathcal D}X
  } \nonumber \\
  &=
  \frac{\int \rho_k(X)  \theta(\lambda_{\rm max}(X) - \lambda_i) \, {\mathcal D}X}{
\int \rho_k(X) \, {\mathcal D}X
  } 
   \frac{\int  \rho_k(X) \, {\mathcal D}X}{
\int \rho_0(X) \, {\mathcal D}X
  } \nonumber \\
  &=
\left \langle  \theta(\lambda_{\rm max}(X) - \lambda_i) \right \rangle_k P_A(\lambda_k|\lambda_0)
\label{eq:At}
\end{align}
This shows that $P_A(\lambda_i | \lambda_0)$ can be computed from multiple ensembles $k$ after unbiasing with $P_A(\lambda_k|\lambda_0)$, i.e.\ $[0^+], [1^+], \ldots, [i^+]$.

As in Eqs.~\ref{eq:WHAMweights} and \ref{eq:WHAMweights2}, we take a weighted average with weights defined by $\alpha_k = \eta_k \times [P_A(\lambda_k | \lambda_0)]^{-1}$ over the range $0 \leq k < i$. Although Eq.~\ref{eq:At} also holds for $k = i$, it reduces to the trivial identity $P_A(\lambda_i | \lambda_0) = P_A(\lambda_i | \lambda_0)$, since $\left \langle  \theta(\lambda_{\rm max}(X) - \lambda_i) \right \rangle_i = 1$. Thus, including $k = i$ does not enhance statistical accuracy and the weighted average is calculated over the range $0 \leq k < i$, resulting in:

\begin{align}
P_A(\lambda_i | \lambda_0) &=
  \frac{ 
  \sum_{k=0}^{i-1} \eta_k
\left \langle
\theta(\lambda_{\rm max}(X) - \lambda_i)
\right \rangle_k 
}{
\sum_{k=0}^{i-1} \eta_k [P_A(\lambda_k|\lambda_0)]^{-1}
} \label{eq:Af}
\end{align}
which matches Eq.~\ref{eq:pp} upon substituting
Eq.~\ref{eq:Q}.

\begin{acknowledgments}
This work was supported by the Research Council of Norway through FRIPRO projects 355155 and 353364, by the FWO (projects G002520N and G094023N), the BOF of Ghent University, and the European Union (ERC Consolidator grant 101086145, PASTIME).
\end{acknowledgments}

\bibliographystyle{apsrev4-1}
\bibliography{biblio.bib}

\end{document}